\documentclass[prd,showpacs,preprintnumbers,floatfix,superscriptaddress,10pt]{revtex4}
\usepackage[mathscr]{eucal}
\usepackage[dvips]{color}
\usepackage[dvips]{graphicx}
\usepackage{epsf}
\usepackage{bm}
\usepackage{amssymb}
\usepackage{amsmath}
\usepackage[normalem]{ulem}
\newcommand{\be}{\begin{equation}}
\newcommand{\ee}{\end{equation}}
\newcommand{\ba}{\begin{eqnarray}}
\newcommand{\ea}{\end{eqnarray}}
\newcommand{\ban}{\begin{eqnarray*}}
\newcommand{\ean}{\end{eqnarray*}}

\newcommand{\bef}{\begin{figure}}
\newcommand{\eef}{\end{figure}}
\newcommand{\bce}{\begin{center}}
\newcommand{\ece}{\end{center}}

\begin{document}

\title{ Jet energy loss in the quark-gluon plasma by stream instabilities}
\author{Massimo~Mannarelli}
\author{Cristina~Manuel}
\author{Sergi Gonz\'alez-Sol\'is}
\affiliation{Instituto de Ciencias del Espacio (IEEC/CSIC),
Campus Universitat Aut\`onoma de Barcelona, Facultat de Ci\`encies, Torre C5 E-08193 Bellaterra (Barcelona), Spain}
\author{Michael~Strickland}
\affiliation{Department of Physics, Gettysburg College, Gettysburg, PA 17325, USA}
\date{\today}
\begin{abstract}
We  study the evolution of the plasma instabilities induced by  two jets of particles propagating in opposite directions and  crossing a thermally equilibrated non-Abelian plasma. 
In order to simplify the analysis we assume that  the two jets of partons  can be described with  uniform distribution functions in coordinate space and by  Gaussian distribution functions in momentum space. 
We find  that while crossing the quark-gluon plasma,  the jets of particles excite  unstable  chromomagnetic and chromoelectric  modes. These fields  interact with the particles (or hard modes) of the plasma  inducing the production of currents; thus,  the energy lost by  the jets is absorbed by both the gauge fields and the hard modes of the plasma.  We compare the outcome of the numerical simulations  with the analytical calculation performed assuming that the jets of particles can be described by a tsunami-like distribution function. We find   qualitative and semi-quantitative agreement between the  results obtained with the two methods.
 
\end{abstract}
\preprint{}
 \maketitle

\section{Introduction}
 The properties of the  matter produced  in heavy-ion collisions  can be studied in several different ways~\cite{Abreu:2007kv}, by observing  electromagnetic signals,  by studying the dissociation of heavy  quarkonia states ($Q \bar Q$) or by studying the propagation of   high $p_T$ partons. The   high $p_T$ partons generated  by hard scatterings in the initial stage of the heavy-ion collision   behave as  hard probes of the produced medium~\cite{Adams:2005dq}. When  a jet of partons travels across the medium it loses  energy  and degrades. The study of the mechanism of energy loss gives interesting information about the properties of the traversed medium (see \cite{Kovner:2003zj} for reviews). The process of absorption of the energy of the jet by the quark-gluon plasma (QGP) has been named jet quenching, because  in the direction of propagation of the jet  one observes a decrease of high energy hadrons, whilst  the number of soft hadrons increases.

Several models have been proposed to describe the jet quenching mechanism, see \cite{Kovner:2003zj}
for a review and  detailed discussions.
In  previous papers~\cite{Mannarelli:2007gi,Mannarelli:2007hj}
we have presented a novel mechanism for describing how the jet loses energy and momentum while traveling in a thermally equilibrated   quark-gluon plasma.  
Since the jet of particles  is  not in thermal equilibrium with the QGP it perturbs and destabilizes the system  inducing the generation of gauge fields.  Some of these gauge modes are unstable and grow  exponentially fast in time, absorbing the kinetic energy of the  jet.  Plasma instabilities in the QGP have been the subject of extensive work in the recent years, see~\cite{Mrowczynski:2006ad} for a review. 
This is so because it was suggested~\cite{Mrowczynski:1994xv,Arnold:2004ti} that the presence of plasma instabilities could help to explain
fast equilibration of the QGP.

In Ref.~\cite{Mannarelli:2007gi,Mannarelli:2007hj} we have studied the propagation  of an energetic jet of particles in an equilibrated quark-gluon plasma  by  using both transport theory and fluid equations~\cite{Manuel:2006hg}. In both cases we have considered many simplifying assumptions.  We have neglected hard modes interactions, assuming that the interaction between the jet and the plasma is only mediated by mean gauge fields. Moreover, the plasma has been considered to be in thermal equilibrium, whereas the jet is assumed to be initially electrically neutral and colorless and described by a uniform distribution function in space and  a tsunami-like distribution in momentum space. 
  
  In the present paper we consider the same simplifying assumptions reported above in a setting where two jets of particles propagate in opposite directions in a thermally equilibrated non-Abelian plasma. We first analytically derive  the growth rate of the unstable gauge field modes and then compare the results with the outcome of numerical simulations. In the numerical simulations we describe the particles of the jet  employing a uniform distribution function in space and  a Gaussian distribution function in momentum space. By varying the width of the Gaussian distribution function we can approximate the tsunami-like distribution function and then compare the results with the outcomes of the analytical study. We find qualitative and semi-quantitative agreement between the results of the two methods. Moreover, the numerical study gives some hints regarding the dynamical processes responsible of the energy loss of the jets. We find that the hard modes of the jet induce the instability of the gauge fields of the plasma, meaning that  the energy associated with the chromomagnetic and chromoelectric  fields  increases exponentially fast in time. In turn, the increase in time of these modes induces  currents  of the particles of the plasma. Therefore, the energy lost by the jets is absorbed by both the gauge fields and the particles of the plasma.  This is a remarkable result, since in our model there are no direct interactions (collisions) among the hard modes of the jets and the hard modes of the plasma. The only interaction that is present is the one mediated by the gauge fields.

It is worth recalling that the interaction of  relativistic streams of particles with (abelian) plasmas  has been studied  in different fields of physics, ranging from inertial confinement fusion,
astrophysics and cosmology. In these experiments one observes that when the particles of the stream are charged, plasma instabilities develop,
leading to an initial stage of fast growth of the electromagnetic fields. These  instabilities have been studied
using a variety of methods, from kinetic theory to hydrodynamics~\cite{Honda,Bret}; experimental evidence of the relativistic
filamentation instability has been reported in Ref.~\cite{exp-inst}.

This paper is organized as follows. In Section~\ref{vlasov} we describe the model and review the basic set of equations. In Section~\ref{analytical} we derive the analytical solutions for the case of two jets described by a tsunami-like distribution function that travel across the equilibrated quark-gluon plasma. Moreover, we derive some general results regarding the energy density of hard-particles.   In Section~\ref{numerical} we present the results of the numerical simulations. We draw our conclusions in Section~\ref{conclusions}.

\section{Fluctuations in the QGP traversed by a neutral jet of particles}
\label{vlasov}

For the sake of simplicity, we consider a plasma consisting only of gluons (adding quarks and anti-quarks is straightforward, as shown in  \cite{Mannarelli:2007gi}) that is initially in a colorless state described by an 
isotropic distribution function. (The treatment  of anisotropic plasmas can be done as in~\cite{Romatschke:2004jh}, where  one assumes that the momentum  of particles is  symmetric only for rotation around an axis.)

For a thermally equilibrated and isotropic  plasma  we use the  Bose-Einstein distribution: 
\be\label{fplasma-iso}
f_{\rm pl} \equiv f^{\rm eq.}_{BE}(p_0) = \frac{1}{e^{p_0/T} - 1} \,, 
\ee
and a uniform distribution function in coordinate space.

Without loss of generality  the particles  that constitute the  jets are taken to be massless quarks
that are  initially in a  colorless state described by a uniform distribution function in coordinate space and a distribution function $ f_{\rm jet} ({\bf p})$ in momentum space.   Several
distribution functions have been proposed \cite{Wang:2001cy} typically expressed in the power-law form
\be
f^n_{\rm jet}(p)\sim \frac{1}{(p^2 + p_0^2)^n}\,,
\ee where  $p_0$  is introduced as an infrared cutoff and $n$ is some positive number. In order to facilitate the analysis of the dynamical properties of the system   in Ref.~\cite{Mannarelli:2007hj} a  tsunami-like distribution  function was used
\be
\label{tsunami}
f_{\rm jet}(p) =  \bar n \;
\delta^{(3)}\left({\bf p}
-  {\bf \Lambda}  \right) \; ,
\ee
that   describes a system of particles of  constant density $\bar n$, all with the same  momentum $\bf p = \Lambda $. In the analytical study of the gauge field instabilities we shall describe the jets by means of this tsunami-like distribution function.

In the numerical study we  use 
a smooth  function of  momentum, which in a certain limit becomes a tsunami-like distribution function. In this way we can compare the results of the numerical simulations with the 
analytical results. More specifically, we  use the  Gaussian distribution
\be
\label{regulated-tsu}
f^\alpha_{\rm jet}({\bf p}) =  N(\alpha) e^{-\alpha(\bf p - \bf \Lambda)^2} \;,
\ee
where $N(\alpha)$ is a normalization factor and $\alpha$ measures the smearing in momentum space. 
In the limit  $\alpha \rightarrow \infty$, one has that $f^\alpha_{\rm jet}({\bf p}) \rightarrow f_{\rm jet}({\bf p})$ and all the particles of the jet have the same momentum $\bf \Lambda$.
We fix the normalization factor by requiring that the particle density is the same for
both the tsunami-like and the Gaussian  distribution functions, namely
\be
\int \frac{d^3p}{(2 \pi)^3} f^\alpha_{\rm jet}(p)=  \int \frac{d^3p}{(2 \pi)^3} f_{\rm jet}(p)\,,
\ee 
which is fulfilled for 
\be\label{normalization}
N(\alpha) = \bar n  (4 \pi \alpha)^{3/2} \;.
\ee

We shall numerically study the system evolution  for various values of $\alpha$ and $\bar n$, in order to  determine under which conditions the plasma instabilities are stronger. It can be  expected from general arguments that plasma instabilities are present for any non-vanishing  value of $\alpha$, see {\it e.g.}  Ref.~\cite{Arnold:2003rq}, however it is important to determine whether the growth rate is sufficiently large so that these instabilities can play a significant role in the evolution of the system.

\subsection{Deviation from equilibrium}

The two jets of particles that propagate  in the non-Abelian plasma excite unstable chromomagnetic and chromoelectric fields destabilizing the system.  In this Section we derive the equations governing the small deviations from    equilibrium of the  distribution functions of  gluons and hard modes. The approach we present below  describes successfully the physics of the soft scales in the plasma, or, the Hard Thermal Loop (HTL) physics as discussed in Ref.~\cite{Bla93,Kel94}.  We  extend that   formalism to add the contribution of the two jets of particles.

Small deviations  from   equilibrium  of the  distribution functions of the gluons of the plasma can be 
described by an  $(N_c^2-1) 
\times (N_c^2-1)$ matrix  in color space $\delta G(p,x)$, where $N_c$ is the number of colors. For the particles of the  jet the  deviation from equilibrium can be described by a $N_c 
\times N_c$ matrix in color space $ \delta W_{\rm jet}(p,x)$.

The  linearized transport equations for the gluons and the particles of the jet can be written as
\be
\label{linear-trans}
p^{\mu}  {\cal D}_{\mu} \delta G(p,x)  =  - g  \: p^{\mu}
  {\cal F}_{\mu \nu}(x) \frac { d  f_{\rm pl}}{d p_\nu} \ ,
\qquad
p^{\mu} D_{\mu} \delta W_{\rm jet}(p,x)  = - g  p^{\mu}  F_{\mu \nu}(x) \frac{\partial f_{\rm jet}}
{\partial p^\nu} 
  \,,
\ee
where the covariant derivatives ${\cal D}_{\mu}$ and $D_{\mu}$   act as
$$
{\cal D}_{\mu} = \partial_{\mu} - ig[{\cal A}_{\mu}(x),...\;]\;,\;\;\;\;\;\;\;
D_{\mu} = \partial_{\mu} - ig[A_{\mu}(x),...\; ]
\;,
$$
with $A_{\mu }=  A^{\mu }_a (x) \tau^a$ and ${\cal A}_{\mu }=  A^{\mu }_a (x) T^a$, and $\tau^a$ and $T^a$ are the $SU(N_c)$ generators in the fundamental and adjoint representations, respectively.
The strength tensor in the fundamental representation is
$F_{\mu\nu}$, while  ${\cal F}_{\mu \nu}$ denotes the field
strength tensor in the adjoint representation.

For massless particles, it is possible to do the following re-definitions
\be\label{redefinition}
\delta G(p,x)  =  - g \frac { d  f_{\rm pl}}{d p_\nu} W^\nu_{\rm pl}(v,x)  \ , \qquad
\delta W_{\rm jet}( p,x)  = -g  \frac{\partial f_{\rm jet}} {\partial p^\nu} W^\nu_{\rm jet}(v,x) \,,
\ee
then considering that the distribution function of the plasma is isotropic one obtains that the transport equations for the gluons of the plasma in the previous equations simplify to
\be
p^{\mu}  {\cal D}_{\mu} \delta G(p,x)  =  - g  \: p^{\mu}
  {\cal F}_{\mu 0}(x) \frac { d  f^{\rm eq.}_{BE}(p_0)}{d p_0} \ , 
\qquad
\delta G(p,x)  =  - g \frac { d  f^{\rm eq.}_{BE}(p_0)}{d p_0} W^0_{\rm pl}(v,x) \;,
\ee
and therefore  $W^\nu_{\rm pl}(v,x) = \delta^{\nu 0} W^0_{\rm pl}(v,x)$. 

Regarding the jet, it is characterized by an anisotropic distribution in momentum space, therefore one has that both $ W^0_{\rm jet}(v,x)$ and $ W^z_{\rm jet}(v,x)$ are non vanishing. Upon substituting the expressions of Eqs.~(\ref{redefinition}) in  Eqs.~(\ref{linear-trans}) one obtains the following equations
\be
\label{lineareq-W}
v^\mu {\cal D}_\mu W^\nu_{\rm pl} = v_\mu {\cal F}^{\mu \nu} \ , \qquad v^\mu D_\mu W^\nu_{\rm jet} = v_\mu F^{\mu \nu} \ ,
\ee
where $v^\mu =  p^\mu/p$ with $p=|\bf p|$.

Since the fluctuations of the distribution function of the plasma and of the jet are not in general color singlets,  it follows that one has to  construct the color currents associated with these fluctuations 

\be
\label{col-current}
\delta j^{\mu }_p(x) = -g \int_p p^\mu \;
 \tau^a {\rm Tr}\big[T^a \delta G(p,x) \big]   \ ,  \qquad
\delta j_{\rm jet}^{\mu }(x) = -\frac{g}{2} \int_p p^\mu \;
 \delta W_{\rm jet}( p,x)  \ ,
\ee
where the momentum measure for massless particles is defined as
\be
\label{measure}
\int_p \cdots \equiv \int \frac{d^4 p}{(2\pi )^3} \:
2 \Theta(p_0) \delta (p^2) \;.
\ee
The color currents generated by the fluctuations of  the distribution functions of the plasma and of the jet enter as  sources  in  the Yang-Mills equation
\be
\label{yang-mills}
D_{\mu} F^{\mu \nu}(x) = \delta j^\nu_t (x) = \delta j_{p}^{\nu}(x) + \delta j_{\rm jet}^{\nu }(x)\; .
\ee

Equations (\ref{yang-mills})  and  Eqs.~(\ref{lineareq-W}) form a set of equations  that has to be solved self-consistently. The 
gauge fields which are solutions of the Yang-Mills equation enter into the transport equations of every particle species and, in turn, affect the evolution of the distribution functions. 

\section{Analytical solution for the Tsunami-like distribution function}\label{analytical}

 Analytical solutions for the color polarization tensor and for the corresponding  dispersion laws of the collective modes  can be obtained when the jet is described by   the tsunami-like distribution function in Eq.~(\ref{tsunami}).
The analysis of the system constituted by a single jet crossing an equilibrated plasma has been done in Ref.~\cite{Mannarelli:2007hj}.  Here we extend that calculation to the case of two anti-parallel jets crossing the plasma.

We briefly recall some results of  the HTL approach and derive the dispersion laws that describe the evolution of  gauge collective modes~\cite{LeBellac}. The dielectric tensor is defined as
\be
\label{dielectric}
 \varepsilon^{ij}(\omega,{\bf k}) = \delta^{ij} + \frac{\Pi^{ij}}{\omega^2}  \,,
\ee
where in order to simplify the notation, we have dropped the color indices.
For the QGP in the HTL approximation one has that
\be
\varepsilon^{ij}_{\rm p}(\omega,{\bf k}) =\Big(\delta^{ij} - \frac{k^ik^j}{k^2} \Big)\varepsilon_{\rm T} (\omega,{\bf k})+  \frac{k^ik^j}{k^2} \varepsilon_{\rm L} (\omega,{\bf k}) \ ,
\ee
where 
\ba
\varepsilon_{\rm L}(\omega,{\bf k}) &=& 1 + \frac{3\omega_{\rm p}^2}{k^2}\left[1+\frac{\omega}{2 k} \log{\frac{\omega -k}{\omega +k}}\right] \ , \\
\varepsilon_{\rm T}(\omega,{\bf k}) &=& 1 - \frac{3\omega_{\rm p}^2}{2k^2}\left[1+\left(\frac{\omega}{2 k} - \frac{k}{2 \omega}\right) \log{\frac{\omega -k}{\omega +k}}\right] \ ,
\ea
are the longitudinal and transverse (with respect to $\bf k$) components of the dielectric tensor of the plasma composed by gluons and massless quark  of $N_F$ flavors,
and $\omega_p^2= \frac 19 g^2T^2 (N_c + N_F/2)$ is the plasma frequency squared.

Now we consider the collective modes which are present in a system described by the tsunami-like distribution function (\ref{tsunami}). In this case one deduces that for the two jets propagating in opposite directions one has that
\ba
\Pi^{ij}_{\rm jet 1}(k) &=& -\omega_{\rm jet}^2 \left( \delta^{ij} + \frac{k^i v^j +k^j v^i}{\omega - {\bf k \cdot v} } -  \frac{(\omega^2 - k^2)v^i v^j }{(\omega - {\bf k \cdot v)^2} }  \right) \ , \\
\Pi^{ij}_{\rm jet 2}(k) &=& -\omega_{\rm jet}^2 \left( \delta^{ij} - \frac{k^i v^j +k^j v^i}{\omega + {\bf k \cdot v} } -  \frac{(\omega^2 - k^2)v^i v^j }{(\omega + {\bf k \cdot v)^2} }  \right) \ ,
\ea
where 
the plasma frequency of the jet is given by $\omega_{\rm jet}^2 = \bar n g^2 /(2\Lambda)$ and the two jets propagate with the same velocity $|\bf v|$  in opposite directions.

\begin{figure}[!th]
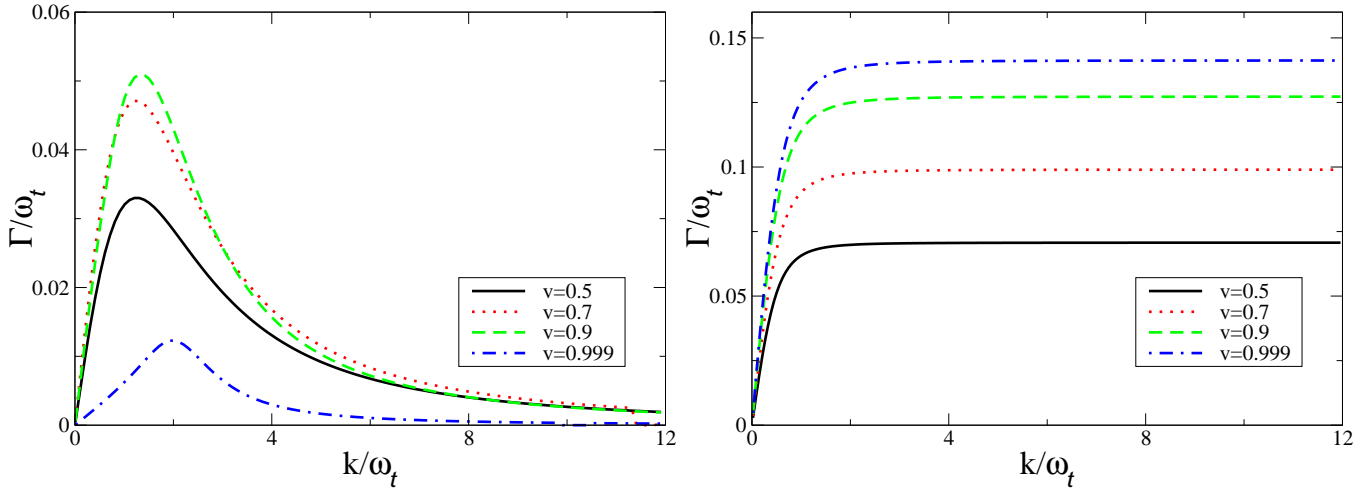

\includegraphics[width=3.5in,angle=-0]{2jets-longitudinal.eps}
\includegraphics[width=3.5in,angle=-0]{2jets-transverse.eps}
\caption{(color online) Imaginary part of the dispersion law of the unstable  modes of the system composed by two jets of particles propagating in opposite directions in a non-Abelian plasma.   The various curves are obtained solving Eq.~(\ref{dispersion-T}) for various values of the velocity and taking  $b=0.01$, see Eq.~(\ref{b}). Left panel: longitudinal component, $\bf k \parallel v$. Right panel: transverse component, $\bf k \perp v$. The full (black) line  corresponds to   $v=0.5$, the dotted (red) line corresponds to $v=0.7$, the dashed (green) line  corresponds to $v=0.9$, and the dot-dashed (blue) line corresponds to  $v=0.999$. These  solutions for the unstable modes have been obtained assuming that the jets of particles have a uniform distribution  in space and the tsunami-like distribution function in momentum space reported in Eq.~(\ref{tsunami}).} \label{fig-analytical}
\end{figure}

In the analysis of the unstable collective modes we  are interested in very short time scales when the Vlasov approximation can be employed. The  effect of the beam of particles is to induce a color current, which provides a contribution to
the polarization tensor. Therefore,  apart from the HTL polarization tensor, we have to include the polarization tensor associated  to the jets. For very short time scales, the polarization tensor of the system is given by the sum of the polarization tensors of its components, meaning that
\be
\Pi^{\mu \nu}(k) = \Pi^{\mu \nu}_{p}(k) + \Pi^{\mu \nu}_{\rm jet 1}(k) +\Pi^{\mu \nu}_{\rm jet 2}(k) \,,
\ee
thus, the dielectric tensor of the system is given by
\be
\label{total-dielectric}
 \varepsilon^{ij}(\omega,{\bf k}) = \delta^{ij} + \frac{\Pi^{ij}}{\omega^2}  \,,
\ee
and the dispersion laws of the collective modes can  be determined solving  the equation
\be
\label{dispersion-T}
 {\rm det}\Big[ {\bf k}^2 \delta^{ij} -k^i  k^j
- \omega^2 \varepsilon^{ij}(k)  \Big]  = 0 \,.
\ee
The solutions of this equation depend on ${|\bf k|}$, ${|\bf v|}$,
$\cos\theta={\bf \hat k \cdot \hat v}$, on
\be
 \omega_t^2 = \omega^2_{\rm p}+\omega^2_{\rm jet} \label{omegat}
\ee 
and on
\be\label{b}
b \equiv \frac{\omega^2_{\rm jet}}{\omega^2_{t}} \ .
\ee
We consider two different cases corresponding to ${\bf k} \parallel {\bf v}$ and ${\bf k} \perp  {\bf v}$ and determine the  behavior of the unstable gauge  modes of the system solving Eq.~(\ref{dispersion-T}).  The results for the imaginary part of the dispersion law  as a function of $k$ for the case ${\bf k} \parallel {\bf v}$ are given in the left panel of Fig~\ref{fig-analytical}. We consider  various values of $v$ and find that the largest instability is obtained for $v\sim 0.9$.
The corresponding results  for the case  ${\bf k} \perp  {\bf v} $ are  reported in the right panel of  Fig~\ref{fig-analytical}. Notice that in both cases  the range of validity of our results is $k < \omega_t$, but we have also reported   the results for larger momenta. In the case  ${\bf k} \perp  {\bf v} $ it happens that for large values of $k$  the growth rate saturates; however, the saturation is achieved in a range of values of the momenta  where the approximations that we are employing are not reliable.

In Table~\ref{tab1} we report the largest value of the growth rate of the longitudinal and transverse components for various values of $b$. Both the longitudinal and transverse growth rates increase with increasing values of $b$. The transverse modes always have a growth rate that is larger than  the one of the longitudinal modes.

\begin{table}[!th]
\centering
\begin{tabular}{|c|c|c|} \hline
$b$ & $\Gamma_{ \parallel}/\omega_t$ & $\Gamma_{\perp}/\omega_t$  \\ \hline
0.01 & 0.02 & 0.12 \\ \hline
0.1 & 0.05  & 0.4 \\ \hline
0.2 & 0.06  & 0.6     \\ \hline
0.5 & 0.08  & 0.9 \\ \hline
\end{tabular}
\caption{Approximate largest value of the imaginary part of the dispersion law  for the parallel and transverse cases, for various values of $b$. For a given value of $b$ the largest value of  $\Gamma_{ \parallel}$ corresponds to $v\simeq 0.8$, whereas for $\Gamma_{\perp}$ it corresponds to $v= 1$.}
\label{tab1}
\end{table}%





\subsection{Energy density of  the hard modes}


The contribution of   the quasiparticles (hard modes) to the energy-momentum tensor of  the plasma  can be determined using the following equation
\be 
{\rm Tr} (\delta j_{\mu \,t} F^{\mu \nu}) = \partial_\mu \Theta^{\mu \nu}_{\rm pl} + \partial_\mu \Theta^{\mu \nu}_{\rm jet} \,,
\ee
where $\delta j_{\mu \,t}$ is the variation of the total current defined in Eq.~(\ref{yang-mills}), and $\Theta^{\mu \nu}_{\rm pl}$ and $\Theta^{\mu \nu}_{\rm jet}$ are the energy-momentum tensors of the plasma and of the jet, respectively.
Then,  the procedure proposed by Blaizot and Iancu  in Ref.~\cite{Blaizot:1994am}  can be used to  single out the energy-momentum tensor for a plasma   close to equilibrium, and  express the result  in terms of $W^0_{\rm pl}$. The same strategy can be generalized for treating  anisotropic, non-equilibrium plasmas as shown in Ref.~\cite{Rebhan:2005re}. 
For the evaluation of the energy density of the plasma one can use the equation
\be
\label{BI-exp} 
{\rm Tr} (\delta j_{\mu \,t} F^{\mu 0}) 
= -\frac{g^2}{2} \int \frac{d^3 p}{(2 \pi)^3} \left\{\frac{d f_{\rm pl} }{d p_\beta} W^\beta_{\rm pl}
(v \cdot D) W_{\rm pl}^0 + \frac{\partial f_{\rm jet} }{\partial p^\beta} W^\beta_{\rm jet}
(v \cdot D) W_{\rm jet}^0 
\right\} \,,
 \ee
and since we are assuming that the plasma is  isotropic, we have that
\be
\label{BI-exp} 
{\rm Tr} (\delta j_{\mu \,t} F^{\mu 0}) 
= -\frac{g^2}{2} \int \frac{d^3 p}{(2 \pi)^3} \left\{\frac{d f^{\rm eq.}_{BE} }{d p_0} W^0_{\rm pl}
(v \cdot D) W_{\rm pl}^0 + \frac{\partial f_{\rm jet} }{\partial p^\beta} W^\beta_{\rm jet}
(v \cdot D) W_{\rm jet}^0 
\right\} \,.
 \ee
Integrating by parts the first term in the equation above  one can identify the energy density associated to the plasma obtaining, in agreement with the results of Blaizot-Iancu, that
\be\label{T00pl}
\Theta^{00}_{\rm pl} = -\frac{g^2}{4} \int \frac{d^3 p}{(2 \pi)^3} \frac{d f_{\rm pl} }{d p_0}
(W^0_{\rm pl})^2  \, .
\ee


In this linear analysis, it is clear that $\Theta^{00}_{\rm pl}$  is always 
positive. However, in Eq.~(\ref{T00pl}) the fluctuations of the gauge fields
have been neglected. When these fluctuations are sufficiently large they can give the dominant contribution to $\Theta^{00}_{\rm pl}$ and  eventually can make this quantity negative. This is indeed what we have observed in the early stage of the numerical simulations that will be presented in the next Section. However, for sufficiently large times the contribution of fluctuations is not dominant and $\Theta^{00}_{\rm pl}$ is positive in agreement with Eq.~(\ref{T00pl}). Regarding  the energy density of the
jet it is not possible to  obtain a close expression analogous to Eq.~(\ref{T00pl}).
Therefore, in order to establish whether the  energy density associated to the jet is positive or negative one has to rely on numerical simulations. 
On general grounds one expects  that if the quasiparticles of the jets are sufficiently energetic   they will lose energy while crossing the plasma. The point is to observe whether the rate at which  the  jet will transfer energy to the plasma will be linear or exponential with time. In the second case  it means that the jet has been able to trigger an instability of the system.

\section{Numerical simulations}\label{numerical}

In this Section we present  the numerical simulations of the real-time evolution of the system composed by the two jets of particles propagating in  an equilibrated and isotropic plasma. For the sake of simplicity we consider the non-Abelian $SU(2)$ gauge group and restrict the gauge field dynamics to $2+1$ dimensions. 

A detailed description of the numerical algorithm that we employ can be found in Ref.~\cite{Rebhan:2005re}. In order to implement the numerical simulation we  introduce  the quantities
\ba
a_{\bf v}&=& - \frac{g^2}{2} \int_{0}^{\infty} \frac{d p }{2 \pi^2} p^2  \frac{\partial f }{\partial p} \label{av} \,,\\
b_{\bf v} &=& - \frac{g^2}{2} \int_{0}^{\infty} \frac{d p }{2 \pi^2} p^2  \frac{\partial f }{\partial p_z} \label{bv}\;,
\ea
 for both  the plasma and the jet particles. Since we are considering an isotropic plasma, one   obtains that  $b_{\bf v}^{\rm pl} = 0$, while $ a_{\bf v}^{\rm pl} = m_D^2 $, where $m_D$ is the Debye mass. 

The two jets propagating in opposite directions  are characterized by the Gaussian distribution function of Eq.~(\ref{regulated-tsu}) with ${\bf\Lambda_1} =(0,0,\Lambda)$ for  the first jet, while ${\bf\Lambda_2} =(0,0,-\Lambda)$ for the other.
For the first  jet   we find  that
\ba 
a_{{\bf v}}^{\rm jet} &=& + g^2  \alpha N(\alpha) \int_{0}^{\infty}  \frac{d p }{2 \pi^2} p^3 e^{-\alpha( p^2 +  \Lambda^2 -2 p v_z \Lambda)}\label{avgauss}\,,\\
b_{\bf v}^{\rm jet} &=& - g^2  \alpha N(\alpha) \Lambda \int_{0}^{\infty}  \frac{d p }{2 \pi^2} p^2 e^{-\alpha( p^2 +  \Lambda^2 -2 p v_z \Lambda)} \label{bvgauss}\;,
\ea

\vspace*{.8cm}\begin{figure}[!th]
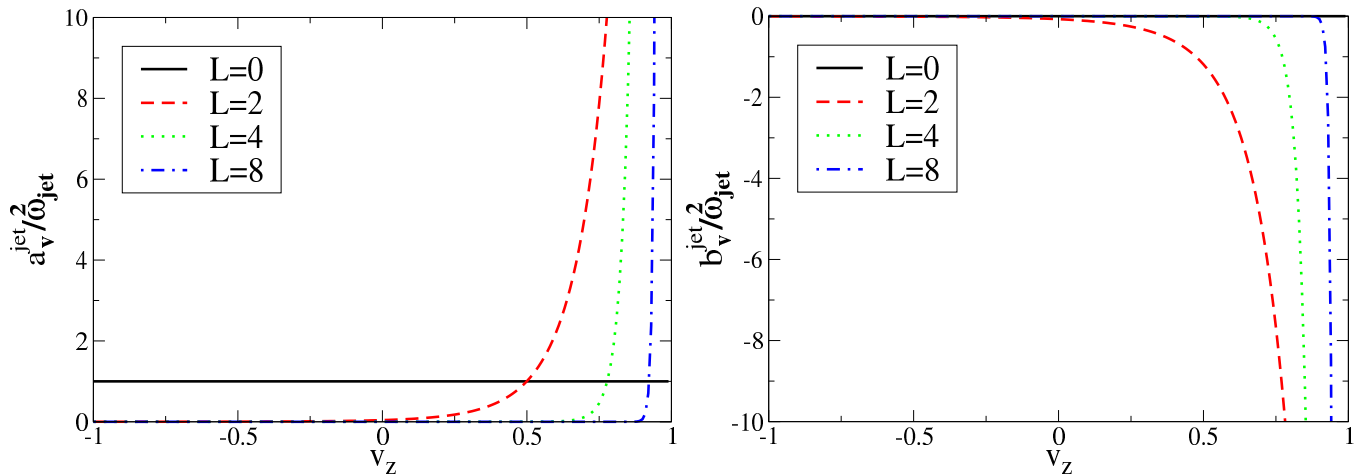

\includegraphics[width=3.5in,angle=-0]{av.eps}
\includegraphics[width=3.5in,angle=-0]{bv.eps}
\caption{(color online) Plot of $a_{{\bf v}}^{\rm jet} /\omega^2_{\rm jet}$ (left panel) and of $b_{{\bf v}}^{\rm jet} /\omega^2_{\rm jet}$ (right panel) defined in Eqs.~(\ref{av-and-bv}) as a function of $v_z$ for four different values of $L$.  With increasing values of $L$ both functions become more and more peaked at $v_z=1$.} \label{fig-of-a}\vspace*{.5cm}
\end{figure}

that can be explicitly evaluated obtaining

\ba \label{av-and-bv}
a_{\bf v}^{\rm jet} &=& \omega^2_{\rm jet} \frac{4 L e^{-L^2}}{\sqrt{\pi}\, {\rm erf}(L)} \left(1 + y_z^2 + y_z \sqrt{\pi}e^{y_z^2}(3/2 + y_z^2)(1 + {\rm erf}(y_z))\right) \,,\\
b_{\bf v}^{\rm jet} &=& - \omega^2_{\rm jet} \frac{4 L^2 e^{-L^2}}{\sqrt{\pi}\, {\rm erf}(L)} \left(y_z + \sqrt{\pi}e^{y_z^2}(1/2 + y_z^2)(1 + {\rm erf}(y_z))\right) \;,
\ea
where ${\rm erf}(x)$ is the error function, where  
\be\label{L} L = \sqrt{\alpha}\Lambda\,, \ee is a dimensionless variable, $y_z=v_z L$ and 
$\omega^2_{\rm jet} = \frac{g^2 \bar n}{2 \Lambda}$
is the plasma frequency of the jet.  (This definition of the plasma frequency is the same we have used in the analytical study.) Notice that 
 $a_{\bf v}$ and $b_{\bf v}$ depend on  $\omega^2_{\rm jet}$ and on functions of $L$  of the form $L^m \exp(-L^2)$, where $m=1,2$.
The color current associated with the jet can now be expressed as 
\be \label{jmu-jet}
\delta j_{\rm jet}^{\mu }(x) = \int \frac{d \bf v}{4 \pi} v^\mu \left(a_{\bf v}^{\rm jet} W^0_{\rm jet}(v,x)+  b_{\bf v}^{\rm jet} W^z_{\rm jet}(v,x) \right) \,,
\ee
where the angular integral can be numerically evaluated  by using the discretization of the unit sphere with
\be\label{discretization}
z_i = -1 + (2 i-1)/N_z, \hspace{.2cm} i=1\dots N_z; \qquad \phi_j=2\pi j/N_\phi, \hspace{.2cm} j=1 \dots N_\phi \,.
\ee
The resulting ``disco balls" are such that they are covered with ${\cal N} = N_\phi \times N_z$ tiles of equal area. In order to test the robustness of the  algorithm  we have performed the integrals increasing the values of 
$N_\phi$ and  $N_z$,  until numerical stability  is achieved.

A plot of the functions $a_{\bf v}^{\rm jet}$ and $b_{\bf v}^{\rm jet}$ is presented in Fig.~\ref{fig-of-a} for four values of $L$. Both  $a_{\bf v}^{\rm jet}$ and $b_{\bf v}^{\rm jet}$ are peaked at $v_z=1$ and the value at the peak increases with increasing $L$. Moreover, the region  where   $a_{\bf v}^{\rm jet}$ and $b_{\bf v}^{\rm jet}$  are  non vanishing becomes narrower with increasing values of $L$ and for $L\to \infty$ they are non zero only for  $v_z=1$ where they diverge. For this reason  one cannot numerically evaluate the integral in Eq.(\ref{jmu-jet}) for arbitrary large values of $L$. Therefore when doing the simulation one cannot consider large vales of $L$; we shall take $L \le 6$.

Notice that $b_{\bf v}^{\rm jet} \le 0$ for any value of $L$ and $v_z$.
For $L=0$ one has that   $a_{\bf v}^{\rm jet}= \omega^2_{\rm jet} $, while $b_{\bf v}^{\rm jet}$ vanishes. The reason is that for $L=0$ there is no flux of particles, the distribution function of the jet becomes  symmetric. The  corresponding expressions for the second jet can be obtained replacing $\Lambda \to - \Lambda$ in the equations above.

An  expression similar to Eq.~(\ref{jmu-jet}) holds for the current of the plasma, however the evaluation of the integral is simpler because we have assumed that  the plasma is  isotropic. Notice that  the  anisotropies in the system are due to the two jets, but the system is still symmetric for rotations around the direction of propagation of the jets, which move in  opposite directions. This means that one can study the gauge field dynamics in $2+1$ dimensions, with the two space dimension  orthogonal to the direction of propagation of the jet.
This allows to  reduce the simulation time.

\subsubsection{Results of the numerical simulation}

\begin{figure}[!th]
\includegraphics[width=4.in,angle=-0]{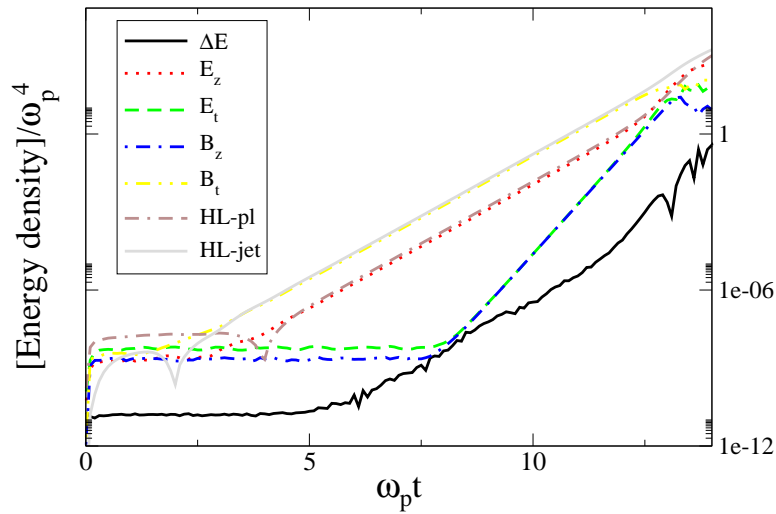}
\caption{(color online) Time  evolution of the absolute value of the energy of longitudinal and transverse   chromomagnetic fields, $B_z$ and $B_t$, of the longitudinal and transverse  chromoelectric fields, $E_z$ and $E_t$, and of the  hard-loop modes of the plasma, HL-pl, and  hard-loop modes of the jet, HL-jet. In this simulation we have used $N_z = 30$ and $N_\phi=30$, see Eq.~(\ref{discretization}), and the parameters $b=0.2$ and $L=4$, see Eqs.~(\ref{b}) and~(\ref{L}).   The full black line  corresponds to $\Delta E$, that is the energy violation measured during the evolution of the system. The results of the numerical simulations become unreliable when $\Delta E $ becomes of the same order of the total energy of the system.} \label{N-series}
\end{figure}

 The simulation depends on several parameters. 
In Fig.~\ref{N-series} we present a typical  plot of the time evolution of the absolute value of the energy of  the  chromomagnetic fields, of the chromoelectric fields,  of the hard-loop modes of the plasma and  of the hard-loop modes of the jet.  The black line corresponds to the violation of the energy measured during the evolution of the system. 

The most unstable gauge field mode corresponds to the transverse chromomagnetic component. 
This is in agreement with the analytical results we obtained in Section~\ref{analytical}  where it was found  that the strongest instability appears in gauge field modes with momentum  orthogonal to the jet velocity. 

However, we can now look more in detail to what happens in the system. The hard modes of the jet lose energy that is absorbed by the gauge fields of the plasma, that exponentially grow. Notice that the growth of the gauge fields of the plasma drives the growth of the hard modes of the plasma. This effect is due to the interaction between the gauge fields and the hard modes of the plasma. In order to understand this point one has to consider that in our simplified model there are no collisions between  particles, therefore  hard particles of the plasma do not directly interact with the hard modes of the jets. The only interaction between the two hard modes is mediated by the gauge fields of the plasma. In other words, what the hard modes of the plasma feel is an increase of the magnetic field and respond by generating a current  and absorbing energy from the gauge fields.

We can schematically say that the sequence of events is the following: The hard modes of the jet induce the instability of the gauge fields of the plasma, meaning that a chromomagnetic field is induced that increases in time. Then, the increase in time of the chromomagnetic field induces a current of hard modes of the particles of the plasma. 

This sequence of events can be deduced/confirmed looking at  Fig.~\ref{N-series}, where one sees that the energy associated to the hard modes of the plasma starts to grow only when the energy of the gauge fields is sufficiently large.

A peculiar effect can be observed in the early stage of the evolution. 
Contrary to what happens at late times, for very short times the hard modes of the plasma lose energy, whereas the hard mode of the jet gain energy. This behavior is due to the initial fluctuations of the hard modes of the plasma and of the jet (that induce fluctuations of the gauge fields as well). However, this effect can be modified changing the initial conditions of the system. As an example if we put fluctuations to zero we find that the energy associated with the hard modes of the plasma is always positive in agreement with  Eq.~(\ref{T00pl}).

Regarding the quantitative analysis, the growth rate of the gauge fields depends on  $b$ and $L$.  In Table~\ref{tab2} we report the values obtained for various values of these quantities.

\begin{table}[htdp]
\centering
\begin{tabular}{|c|c|c|c|} \hline
$b $ & $L=2$ & $L=4$ & $L=6$  \\ \hline
0.1 & 0.09  & 0.21  &  0.24  \\ \hline
 0.2 & 0.12 & 0.29 & 0.31 \\ \hline
0.5 &  0.15 & 0.32  &   0.35     \\ \hline
\end{tabular}
\caption{Values of the growth rate, $\Gamma/\omega_t$, of the unstable gauge fields for various values of $b$ and $L$.}\label{tab2}
\end{table}%

The results of the numerical simulation are in qualitative and semi-quantitative agreement with the analytical results. In both cases the growth rate of the most unstable modes correspond to  the transverse ones and the growth rate increases with increasing values of $b$. Moreover, the values of the growth rate in the two approaches are quite similar as can be seen comparing Table~\ref{tab1} with Table~\ref{tab2}. This is a remarkable result, because  in the numerical simulation one has  an {\it integrated} growth rate, meaning that  modes with all momentum compatible with the lattice cutoff contribute to the  growth rate. On the other hand, the growth rate of the analytical case  has been determined considering the largest value of the curve in Fig.~\ref{fig-analytical}.

\section{Conclusions}\label{conclusions}

We have studied a system composed by an equilibrated and isotropic quark-gluon plasma traversed by two jets of particles propagating in opposite directions. We have analyzed the instabilities of this system in two different ways. First, we have 
carried out an analytical study using a very simplified description of the  distribution function of the jets, the so-called tsunami-like distribution, while the plasma has been  described by a Bose distribution function.  Then, we have performed a numerical study, assuming that the distribution functions of the jets  are two Gaussian distribution in momentum. In both cases we have assumed that the plasma and the jets have  uniform distribution in coordinate space. In order to simplify the  numerical analysis we have restricted the gauge field dynamics to  $2+1$ dimensions and we have considered  the $SU(2)$ gauge symmetry.

Comparing the results of the two approaches we have found qualitative and semi-quantitative agreement for the behavior of the unstable gauge fields. In both cases  the most unstable modes correspond to the transverse chromoelectric and chromomagnetic fields. 
Moreover,  in both approaches we find that the growth rate  is about an  order of magnitude smaller than the total plasma frequency $\omega_t$, defined  Eq.(\ref{omegat}), and increases with increasing values of $b$, see Eq.(\ref{b}). The plasma instabilities fully develop on time scales of the
order of $t \sim 10/\omega_t$ and we can estimate the upper bound of this time scale   evaluating the plasma frequency in a weak coupling scenario. Considering a temperature $T \sim 300$ MeV we find that $t\sim 1$ fm/c. 

From the numerical simulations one realizes that the energy lost by the hard modes of the jets is absorbed by both the gauge fields and the hard modes of the plasma.    
It happens that the hard modes of the jet induce the instability of the gauge fields of the plasma and therefore  the energy associated with the chromomagnetic and chromoelectric  fields of the plasma  increases exponentially fast in time. Then, the increasing  chromomagnetic and chromoelectric  fields  induce  currents  of the particles  of the plasma and  transfer energy to these hard modes.

The analysis we have performed is simplistic in many ways. In particular, we have assumed that both the plasma and the jets have a uniform distribution in space. 
It should be possible to improve the treatment presented here in future studies.  One should localize the jets in space-time, with the probable associated effect of having a localized
energy deposition. So far, we have only considered the very short time evolution of the system.
At sufficiently long times, radiative and collisional  processes should be included, with the
addition of collision terms in the transport equation. In a weak
coupling scenario, the collisional times are suppressed by a factor $1/g^2$ with respect to the time
of  development of the instabilities. Collisions will probably stop the growth of the gauge fields, but  they will represent a later mechanism of jet energy loss. It might be interesting to see if one could
incorporate in a numerical code all the above mentioned effects, along the line presented in 
Ref.~\cite{Qin:2007rn}

 Even with the simplifications employed in this paper we believe that    the effect  we discuss, which
is based on collective plasma phenomena,
 might be  relevant for the description of the energy loss mechanism of relativistic jets traversing a quark-gluon
plasma. 

\begin{acknowledgments}
This work has been supported by the Spanish grant
FPA2007-60275. 
\end{acknowledgments}

\end{document}